# Variability of extragalactic X-ray jets on kiloparsec scales

Eileen T. Meyer [1] ✉, Aamil Shaik[1], Yanbo Tang[2], Nancy Reid[3], Karthik Reddy[1,4], Peter Breiding[5], Markos Georganopoulos[1], Marco Chiaberge[5,6], Eric Perlman[7], Devon Clautice[7], William Sparks[8,9], Nat DeNigris[1,10] & Max Trevor[1,11]

Unexpectedly strong X-ray emission from extragalactic radio jets on kiloparsec scales has been one of the major discoveries of Chandra, the only X-ray observatory capable of sub-arcsecond-scale imaging. The origin of this X-ray emission, which appears as a second spectral component from that of the radio emission, has been debated for over two decades. The most commonly assumed mechanism is inverse-Compton upscattering of the cosmic microwave background by very low-energy electrons in a still highly relativistic jet. Under this mechanism, no variability in the X-ray emission is expected. Here we report the detection of X-ray variability in the large-scale jet population, using a novel statistical analysis of 53 jets with multiple Chandra observations. Taken as a population, we find that the distribution of $P$ values from a Poisson model is strongly inconsistent with steady emission, with a global $P$ value of $1.96 \times 10^{-4}$ under a Kolmogorov–Smirnov test against the expected uniform (0, 1) distribution. These results strongly imply that the dominant mechanism of X-ray production in kiloparsec-scale jets is synchrotron emission by a second population of electrons reaching multi-teraelectronvolt energies. X-ray variability on the timescale of months to a few years implies extremely small emitting volumes much smaller than the cross-section of the jet.

Enabled by its unrivalled spatial resolution, one of the major discoveries of the Chandra X-ray Observatory has been the detection of hundreds of resolved kiloparsec-scale jets from supermassive black holes[1]. Chandra has discovered far more of these than initially thought likely, due to the fact that many show an anomalously high and hard X-ray flux from a second spectral component, which was previously unknown. Shortly after the first discovery of the X-ray jet in PKS 0637-752[2,3], the inverse-Compton upscattering of the cosmic microwave background (IC/CMB) model was proposed to explain the unexpectedly high X-ray emission[4,5], and until recently it was by far the preferred interpretation. The IC/CMB model requires that the jets remain highly relativistic on kiloparsec scales (bulk Lorentz factors $\Gamma \approx 10$), and also very well aligned to our line of sight. The relativistic particles in the jet unavoidably inverse-Compton upscatter the microwave-energy cosmic microwave background (CMB) photons to higher energies, and with large $\Gamma$ as well as a jet aligned at a small angle, the Doppler boosting of this emission can be extreme enough to explain the observed X-ray fluxes. In this scenario the electrons responsible for upscattering the

[1]Department of Physics, University of Maryland, Baltimore County, Baltimore, MD, USA. [2]Department of Mathematics, Imperial College London, London, UK. [3]Department of Statistical Sciences, University of Toronto, Toronto, Ontario, Canada. [4]School of Earth and Space Exploration, Arizona State University, Tempe, AZ, USA. [5]The William H. Miller III Department of Physics and Astronomy, Johns Hopkins University, Baltimore, MD, USA. [6]Space Telescope Science Institute for the European Space Agency, Baltimore, MD, USA. [7]Department of Aerospace, Physics and Space Sciences, Florida Institute of Technology, Melbourne, FL, USA. [8]SETI Institute, Mountain View, CA, USA. [9]Space Telescope Science Institute, Baltimore, MD, USA. [10]Department of Astronomy, University of Massachusetts Amherst, Amherst, MA, USA. [11]Department of Physics, University of Maryland, College Park, College Park, MD, USA. ✉e-mail: meyer@umbc.edu

CMB to X-ray frequencies are very low in energy. Because of the very long cooling lifetimes of such electrons and the absolute steadiness of the CMB, any observed variability in the X-rays is incompatible with the IC/CMB model.

Motivated by previous reports of month- and year-timescale variability in the X-ray jet of Pictor A[6–8], we undertook an archival analysis of all multiple-observation kiloparsec-scale X-ray jets. Our analysis sample comprises nearly all known X-ray jets imaged more than once by the Chandra Advanced CCD Imaging Spectrometer (ACIS) instrument, a total of 53 sources. The average number of observations per source in our sample is 3.4, with a mean spacing of 2.6 yr. More than half (30) have been observed only twice, while a small number (6) have more than five distinct observations. We have excluded two sources (3C 305 and 3C 171) where the X-ray emission associated with the radio jet has been attributed to jet-driven gas[9,10], as well as the two most deeply observed jets in the known population, M87 and Centaurus A. The X-ray emission in these latter sources is consistent with the falling tail of the radio-optical synchrotron spectrum[11–13]. Both M87 and Cen A have been far more deeply and frequently observed than typical of the remaining sample (with 50 and 43 distinct observations, respectively), and this is probably one of the reasons that X-ray variability has been reported in both cases[6,7,14]. With the exception of Pic A, variability has not been reported for any other source in our sample of 53. Previously, a knot in the jet of Pic A was seen to fade over a timescale of a few years[6] with a reported significance of 3.4$\sigma$, and potential low-level variability has been reported for the terminal hotspot[7]. Unlike the jets of M87 and Cen A, however, the X-ray emission in Pic A is clearly from a second emission component[15], distinct from the radio-optical, and is thus retained in our sample.

For each X-ray jet, we produced a deep co-added X-ray image by aligning and stacking all available epochs. We obtained complimentary radio-wavelength images at matching or better resolution, predominantly using archival data from the Very Large Array (Supplementary Table 1). We aligned and overlaid the radio and X-ray images of the jet to identify distinct compact zones of emission, referred to as knots (Fig. 1). At these scales, individual knots are causally independent (the minimum knot-to-knot spacing among all jets is 650 light yr deprojected, and on average it is nearly 200,000 light yr). Our analysis assesses variability for individual knot regions rather than the entire extended jet due to the potential for independent and random variations along the jet to cancel out and reduce the statistical power of our test.

For each knot we define a consistent geometric area (knot region) to extract the total X-ray counts from each epoch, as depicted for the source PKS 1136-135 in Fig. 1. A corresponding larger region is used to estimate the counts from background emission. Further details of the data analysis and preparation are given in Methods and Supplementary Section 1, and a full description of the source and region properties and measurements are given in Supplementary Tables 2–5.

Our likelihood function is a straightforward application of Poisson statistics (Methods), with a null hypothesis of a steady source rate for each individual knot region, taking into account a varying background. For each knot, the observed source and background counts for all individual observations (epochs), along with the corresponding exposure (in units of area × time) in the respective regions for that epoch, are the data. We compute for each knot a $P$ value for the test of the null hypothesis of a steady source rate, and a maximum-likelihood estimate of the mean count rate for the knot, $\bar{\mu}$. A low $P$ value means that the null hypothesis can be rejected in favour of an intrinsically variable source rate. Importantly, for a population of non-variable X-ray sources the single-region $P$ values should follow a $U(0, 1)$ distribution.

Using the traditional threshold of $P < 0.05$ to reject the null hypothesis, we find that 18 regions out of 155 tested are inconsistent with a steady source rate (compared with ~8 expected), and the distribution shows a notable excess at low $P$ values (Extended Data Fig. 1). A comparison of the 155 single-region $P$ values with a $U(0, 1)$ distribution using a one-sided Kolmogorov–Smirnov (KS) test confirms the excess of low $P$ values, with a global KS $P$ value for the comparison with $U(0, 1)$ of 0.000196. This clearly indicates that the observations are not consistent with a universal constant source rate for the entire population. (To avoid confusion, we hereafter refer to the $P$ values from the maximum-likelihood function for a single region as 'single-region' $P$ values, and the $P$ values associated with the KS test of the entire distribution of the latter as 'global' or 'KS-test' $P$ values.)

The departure from a $U(0, 1)$ distribution clearly indicates some amount of variability, but does not immediately yield constraints on the fraction of the sample that is variable (if less than unity) or the typical scale of the variability. It is theoretically possible for the population of X-ray emitting jets to be a mixture of variable and non-variable X-ray emitters, and some degeneracy to exist between the fraction of the sample that is intrinsically variable and the amplitude of that variability.

To better constrain the fraction of the sample displaying variability and its characteristic scale, we generated simulated data sets differing in these two respects, and then compared the resulting $P$-value distributions to look for those most like our real data. The simulated data were generated using the real average exposure values for our knot regions as well as their estimated mean count rates $\bar{\mu}$. We used a simple model of variability, of two epochs with a fixed percentage variance relative to mean, and used a KS test to compare a simulated single-region $P$-value distribution ($n = 10,000$) with our real distribution. The resulting global $P$ values from the KS-test comparisons are displayed in Fig. 2 as a two-dimensional colour map with axes of variability amplitude (percentage relative to mean rate $\bar{\mu}$) against the fraction of the sample that is variable versus steady. Lighter colours (indicating simulation/data KS-test $P$ values closer to 1) show the simulations most resembling the data. While there is notable degeneracy between characteristic amplitude and how much of the sample is variable, the data are best matched by simulations in which between 30 and 100% of the sample is variable.

Variability in the X-ray emission from large-scale jets on timescales of a few months to years is not compatible with the IC/CMB mechanism: the CMB is completely steady, and the electrons upscattering it are very low energy (Lorentz factor $\gamma \approx 100$–$1,000$), with extremely long cooling timescales, many orders of magnitude longer than the light-crossing time for the jet[6,16]. However, a synchrotron origin is compatible with short-timescale variability[6], as the X-rays must be emitted by very energetic multi-teraelectronvolt electrons, where the cooling timescale is accordingly much shorter. Combined with a very small emitting volume (of the order of light months), it is possible to produce the observed variability. We therefore suggest that this variability is strong evidence that the X-ray emission in most jets is more likely to be synchrotron radiation from a very high-energy electron population, which in turn implies lower power requirements[17]. It is notable that the required small volumes are, however, in conflict with the typical assumption of particle acceleration distributed throughout the jet cross-section[18] and present a serious challenge to current theoretical models.

Our sample exhibits considerable variety in terms of jet and knot region characteristics such as jet orientation, length, redshift, knot distance from the central engine, spectral type of the host nucleus (broad or narrow/absent optical emission lines) and so on. The sample also covers several orders of magnitude in jet power (as proxied by the low-frequency radio luminosity). Because of the poor statistics at the individual-region level, very little can be learned from looking at individual jets or regions with low $P$ value. Instead we adopt a population approach to look for the influence of jet or region properties on variability. To test if any particular source characteristics were more associated with variability than others, we split the sample into two subsets and reran the one-sided KS test of the single-region $P$-value distribution against a $U(0, 1)$ distribution for each subset. The results are summarized in Table 1, where we give the selected characteristic, the subsample definitions and size, and the resulting KS-test $P$ values.

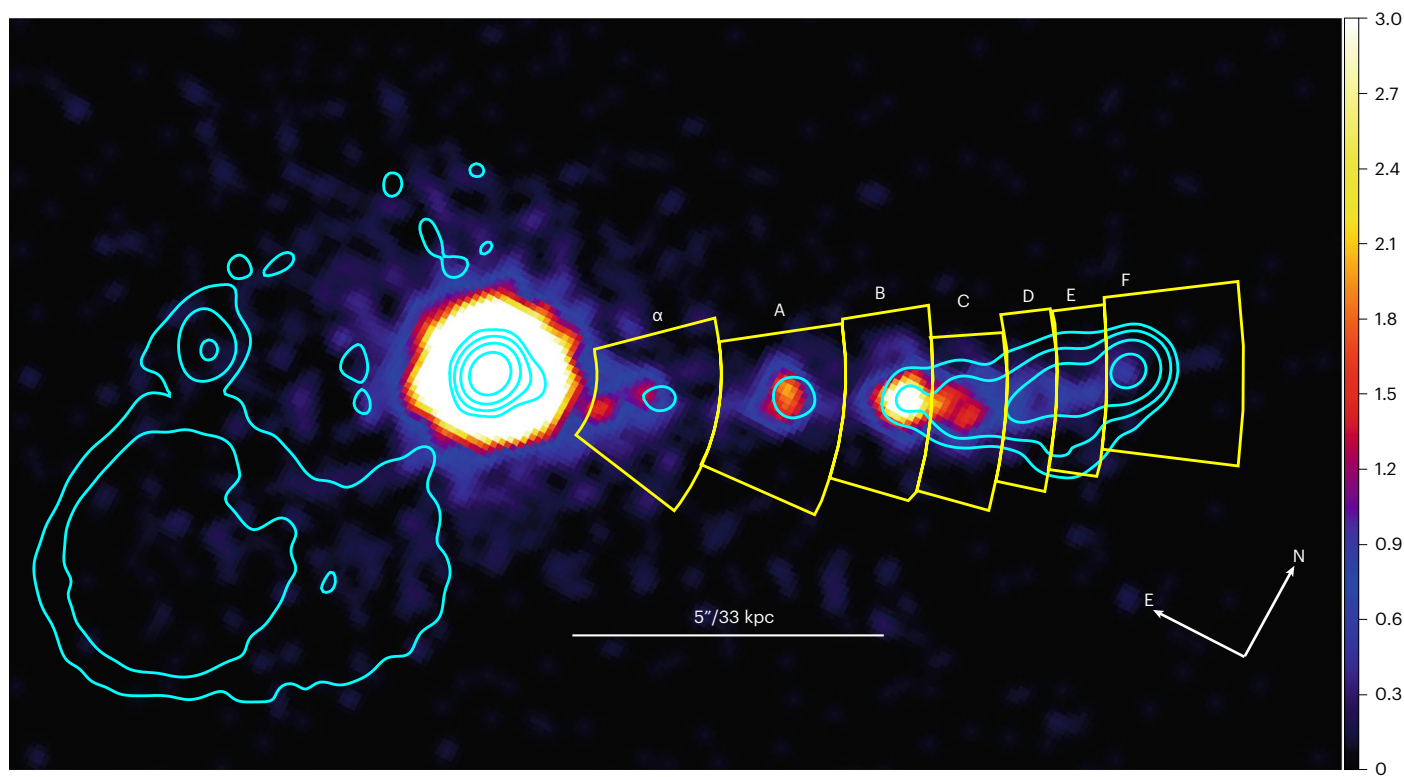

**Fig. 1 | A Chandra X-ray image (0.3–7 keV) of PKS 1136-135, one of the 53 X-ray emitting jets in our sample.** The colour scale is in counts and the image has been lightly smoothed (Gaussian kernel of width 3 pixels). The brightest region, known as the core of the jet, is the location of the galaxy centre and black hole. Overlaid in cyan are contour lines from a radio image at 5 GHz (the counterjet radio lobe is visible to the left of the core). The regions outlined in yellow and labelled are individual emitting regions (knots) identified through a cross-comparison of the radio and X-ray structure.

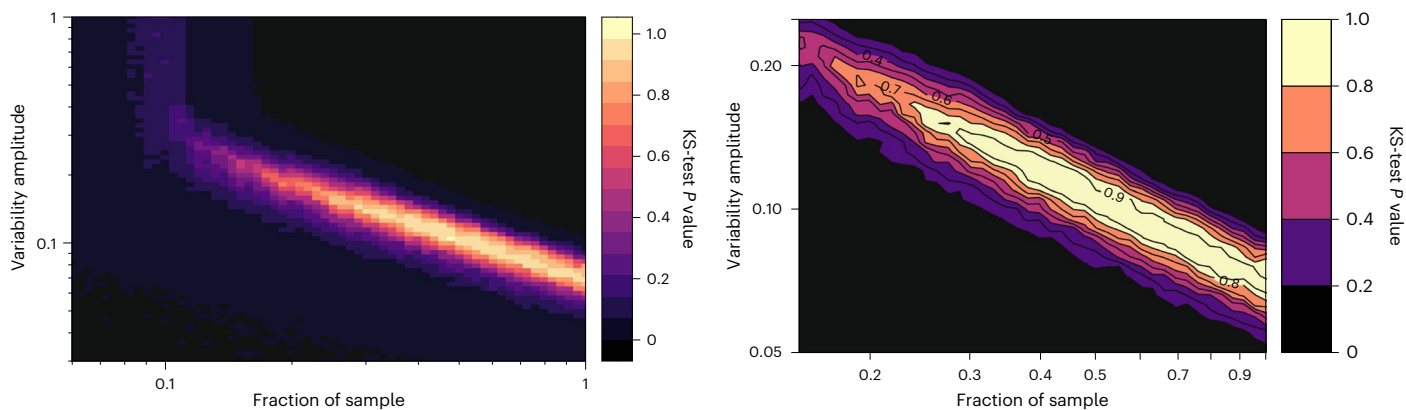

**Fig. 2 | Level plots showing the $P$ value for a one-sided KS test comparing simulated data sets with our observed single-region $P$-value distribution.** Here we have varied the simulated fraction of the population that is variable ($x$ axis) and the amplitude of the variability for that subset ($y$ axis). Higher KS-test $P$ values (lighter colours) indicate closer agreement with the observed data. On the left we show the full results of all simulations, while on the right we show a filled contour plot corresponding to a subset of most probable parameters. While there is a degeneracy between the amplitude of variability and the variable fraction, the simulations imply that at least 30% of the population is variable and that the typical scale of the variability is of the order of 10% compared with the mean.

For subsets where the KS-test $P$ value is lower than that of the full sample, the final column in these tables gives a percentage, which is the percentage probability of obtaining the lower KS-test $P$ value purely by chance for a subselection of size $n$ from the full parent population (that is, the percentile). A histogram depicting the $P$-value distribution for one of the split sample analyses is shown in Fig. 3.

We interpret the combination of a lowered global $P$ value with a low percentile for one subset to mean that the compliment subset either contains more steady jets relative to the subset with a lowered global $P$ value, or preferentially displays variability below our detection sensitivity for those regions, which varies due to factors such as the relative level of background radiation and intrinsic brightness. We found that 7 subselections (out of 12 in total) yield a sample with a lower KS-test $P$ value than the full population, with percentiles ranging from 0.21 to 10.14%. These cases are briefly discussed below, with additional discussion in Supplementary Section 1.

Table 1 | Results of the one-sided subsample KS tests against the $U(0,1)$ distribution, not adjusted for multiple comparison

| Characteristic | Subset | n | Break | KS-test P value | Percentile |
|---|---|---|---|---|---|
| Jet angle | High | 78 | $\theta \geq 15.5°$ | 0.0001212 | 6.17 |
| | Low | 77 | $\theta < 15.5°$ | 0.1227422 | |
| Average source counts | Many | 89 | $n \geq 40$ | 0.0019570 | |
| | Few | 66 | $n < 40$ | 0.0467632 | |
| Background/source count ratio | Low | 66 | $\langle n_{src}/n_{bg} \rangle < 0.1$ | 0.0010641 | |
| | High | 89 | $\langle n_{src}/n_{bg} \rangle \geq 0.1$ | 0.0246957 | |
| Core dist. | Far | 83 | $d \geq 5''$ | 0.0000327 | 2.22 |
| | Near | 72 | $d < 5''$ | 0.2920501 | |
| Jet length | Long | 82 | $l \geq 10''$ | 0.0008115 | |
| | Short | 73 | $l < 10''$ | 0.0431792 | |
| Knot sequence | Later knots | 102 | — | 0.0023121 | |
| | First knots | 53 | — | 0.0208292 | |
| Knot type | Non-hotspot | 110 | — | 0.0025399 | |
| | Hotspot | 45 | — | 0.0076676 | |
| Radio power | Low power | 60 | $L_{300 MHz} < 10^{43}$ erg s$^{-1}$ | 0.0000654 | 2.34 |
| | High power | 95 | $L_{300 MHz} \geq 10^{43}$ erg s$^{-1}$ | 0.1019381 | |
| Spectral type | Broad lined | 120 | — | 0.0000934 | 10.14 |
| | Narrow lined | 33 | — | 0.4180435 | |
| Redshift | Low | 75 | $z < 0.6$ | 0.0000035 | 0.36 |
| | High | 80 | $z \geq 0.6$ | 0.2296055 | |
| Jet class | Not CDQ | 81 | — | 0.0001501 | 7.33 |
| | CDQ | 74 | — | 0.1272601 | |
| Jet class + redshifts | Not high-z CDQ | 100 | $z < 0.6$ or not CDQ | 0.0000032 | 0.21 |
| | High-z CDQ | 55 | $z \geq 0.6$ and CDQ | 0.7034346 | |

CDQ, core-dominated quasar.

A lower KS-test P value is found for the subsample comprised of regions more than 5″ from the bright central core. This is consistent with expectations, as the simulation summarized in Fig. 2 as well as a simple histogram of the percentage variation from mean for all regions (Extended Data Fig. 2) strongly suggests that the typical variability level in the population is at the few to tens of per cent level (as opposed to large orders-of-magnitude flares). Variability at this level may not be detectable in some test regions due to shorter exposures, intrinsically low source flux, relatively high levels of background or any combination of these factors, which together set a lower limit on the amplitude of variability we can detect. We call this limit the sensitivity threshold, and for high-background regions (common for knots near the bright base of the jet or core) the sensitivity threshold may be well above the typical source variability amplitude, while this is less likely to be the case for regions with low background or encompassing relatively bright knots. A further discussion of the sensitivity threshold and its estimation is given in Supplementary Section 2.

In the remaining cases of subsets with a lower KS-test P value, the compliment subset (with a higher P value) consists of source populations where the likelihood of steady IC/CMB emission being enhanced or dominant is supported on theoretical grounds. For all such 'excluded' populations we cannot reject the null hypothesis that the population is non-variable in the X-rays. In particular, we find that higher-redshift jets are consistent with the steady-flux hypothesis, while removing them increases the indication of variability (via decreased global KS-test score) in the remaining low-redshift sample. Because the CMB energy density increases markedly with redshift as $(1 + z)^4$, there has long been an expectation that the IC/CMB mechanism of X-ray production would increasingly dominate with redshift, leading to a strong correlation between X-ray/radio flux ratio and redshift in resolved jets[19] and an enhancement in the total X-ray emission even for unresolved jets[20]. Until recently, the severe lack of high-redshift sources among the known X-ray jets (with only a handful above $z \approx 3$) made it difficult to determine whether the lack of an observed correlation was a contradiction to this expectation or merely a reflection of wide scatter in the intrinsic X-ray/radio flux ratio and/or inadequate sampling[21,22]. In recent years, however, several very high-redshift ($z > 4$) X-ray jets and jet candidates have been identified with high X-ray/radio flux ratios strongly suggesting an IC/CMB origin[23–25]. Studies of high-redshift quasars also show marked (>50%) total X-ray flux enhancements at high redshifts ($z > 3$), which has been attributed to IC/CMB emission in unresolved large-scale jets[20]. Our results, alongside these recent discoveries, suggest a gradual change in the jet population, from synchrotron dominated at lower redshift, giving way to increasing IC/CMB dominance in the population at high redshift, presumably due to the enhancement of the CMB. While a small handful of jets at lower redshifts, under favourable conditions, may have X-ray emission produced by IC/CMB[26], recent observations with the Fermi gamma-ray observatory also generally disfavour an IC/CMB origin for many low- to moderate-redshift jets[27,28].

In addition to those at low redshift, we also find that low-power jets (300 MHz power, $L_{300 MHz} < 10,643$ erg s$^{-1}$) are more discrepant from the $U(0,1)$ population (percentile 2.34%) than the full sample, while we cannot exclude the null hypothesis for the high-power sources. Because the IC/CMB X-ray model generally requires jets with highly relativistic ($\Gamma \approx 10$) speeds on kiloparsec scales, a possible interpretation is that IC/CMB X-rays are more often produced in powerful, highly collimated jets. This is in keeping with general observations that jets with an anomalous X-ray component are mostly hosted by powerful quasars, though a more careful statistical analysis with complete samples is needed to confirm this.

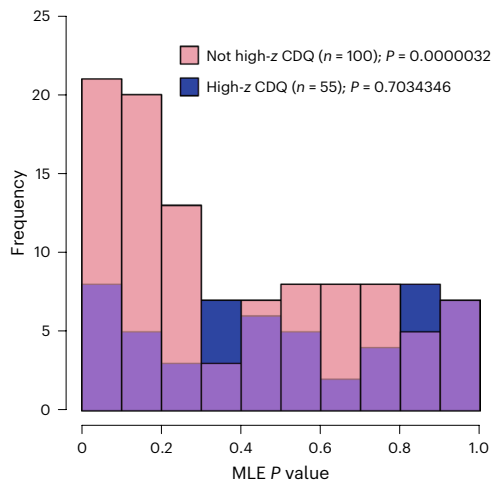

**Fig. 3 | Histogram of the single-region *P* values from the directional test, not adjusted for multiple comparison.** In pink, the subset of sources that excludes core-dominated quasars at high redshift shows a single-region *P*-value distribution more discrepant from a $U(0, 1)$ distribution than the full sample. This implies that high-redshift core-dominated quasars (blue) could represent a true non-variable population, in keeping with theoretical expectations.

Two additional characteristics, the 'jet class' and orientation angle, give a subset with a much lower global KS-test *P* value than the full sample. Here the subpopulations that conversely appear consistent with the steady-flux hypothesis are core-dominated quasars and (with notable overlap of ~90%) sources with a small estimated angle to the line of sight. This is consistent with theoretical expectations, as inverse-Compton emission is more sensitive to the increased Doppler boosting of the relativistic jet with decreasing angle to the line of sight[29,30]. When we make a double selection, excluding only the 47 high-redshift core-dominated quasars, the one-sided KS-test *P* value against a $U(0, 1)$ distribution of the remaining sample decreases to 0.0000032, with a 0.21% probability of occurring by chance. Sources classed as core-dominated quasars are usually broad lined and are assumed to have jets oriented at small angles to the line of sight. Here the orientation angle appears to be the critical factor, since we find that a subselection of broad-lined sources from our sample shows these sources to have a similar 'variability signal' in terms of the global KS-test *P* value ($9 \times 10^{-5}$, with 10% possibility of being drawn by chance). A further discussion of the subsample analysis results and their interpretation is contained in Supplementary Section 1.

In addition to the results presented here, two nearby and very well known X-ray jets have been extensively and deeply studied with the Chandra observatory, more so than any other jets. These are Cen A and M87. In comparison to our sample, these low-redshift jets have far more extensive and deeper observations, with 43 and 50 distinct observations totalling 1,129 and 460 ks of total time on source, respectively (the median total time for our sample is in comparison 63 ks). Previous works have already extensively described their characteristics, including the unprecedented (and unrepeated) X-ray flare from knot HST-1 in M87, where the X-ray flux increased by a factor of over 50 during the mid-2000s, before fading back to its previous level[31]. The variability in Cen A, by contrast, appears as a gradual dimming consistent with adiabatic expansion[32]. In terms of spectrum, the unusually deep observations of M87 and Cen A make it clear that the X-ray emission is from the high-energy falling tail of the radio synchrotron component, and the X-ray variability is thus consistent with this picture. Since we are mainly interested in testing the inverse-Compton scenario for jets in which the X-rays are or might be from a second emission component, we do not include these sources in our analysis. The one other jet previously reported as variable is Pic A. In contrast to M87 and Cen A, the X-ray emission is clearly from a distinct second spectral component[15], and previous works have suggested variability in both the jet and hotspot[6,7]. In this work we confirm that Pic A is strongly variable and find that five of the nine regions in the jet, including the hotspot, are variable at the $P < 0.05$ level, and the mean of the absolute value of the percentage variation from $\bar{\mu}$ is 18%. The fact that all three of the most deeply observed X-ray jets show variability lends support to the population-based findings in this work.

For completeness and as a check of our methods, we also analysed the two X-ray jets previously excluded due to the X-ray emission being from surrounding gas (3C 305 and 3C 171). In these cases no variability is expected due to the scale of the X-ray emission. Indeed, in our analysis none of the regions give single-region *P* values less than 0.18 and thus all are consistent with a steady source rate.

The short timescales of X-ray variability observed in this and previous works implies that the emitting regions responsible are much smaller that the width of the jet. By the light-crossing-time argument a flare event on the order of a year cannot occur in a region larger than a few parsecs, vastly smaller than the typical resolved cross-section in the radio which is of the order of a kiloparsec or more. This requires extremely localized particle acceleration, more easily achieved through magnetic reconnection than through the usually assumed shock acceleration[33]. At present we lack a clear understanding of particle acceleration in extragalactic jets. Future theoretical work must take into account the requirements of short-timescale X-ray variability and the implied multi-teraelectronvolt-energy electrons produced on scales of hundreds of thousands of light years from the central engine.

## Methods
### Initial Chandra data analysis and imaging
The initial list of known X-ray jets was taken from the XJET online database[34] and from a search of the literature, a total of 199 jets. A comprehensive catalogue of known X-ray jets to date can be found in the recent ATLAS-X publication[1]. For the 55 multiply observed sources in our sample, we downloaded all available observations taken with Chandra's ACIS[35]. Supplementary Table 2 lists the names and properties of the sample of 55 jets, and a complete list of the Chandra observations used can be found in Supplementary Table 3.

For the X-ray data analysis, we used CIAO[36] 4.12 and CALDB (https://heasarc.gsfc.nasa.gov/docs/heasarc/caldb/caldb_intro.html) 4.9.2. Each observation was first reprocessed using the chandra_repro command. We applied a standard energy filter including only events from 0.4–8 keV. Excluding all bright sources in the field of view, we extracted the background counts and used the lc_clean task to examine the data for background flares with a good time interval bin time of 259.28 s and cutting intervals with rates more than $2\sigma$ from the mean count rate. The total exposure times before and after filtering are given in Supplementary Table 3.

We used these filtered data to produce initial subpixel images centred on the core of the X-ray jet for each observation, downsizing by a factor of 5 (a pixel scale of 0.0984″). We then co-added all epochs to make a deep subpixel image for comparison with the radio imaging and to identify knot regions. Before co-adding, we corrected for pointing offsets by aligning the point-source-like jet core in all exposures. To do so we manually calculated the offsets between the earliest available observation and all subsequent observations using the centroid task in CIAO. Using these offsets, we corrected the aspect solution and reprojected all observations to match the reference epoch. We then repeated our filtering and subpixel binning process on the reprojected data to create a reprojected subpixel image for each epoch. These were finally merged using dmmerge to produce the deepest combined subpixel image of the X-ray jet for morphological study. For sources with observations in both the FAINT and VFAINT modes, which cannot be merged using CIAO's dmmerge command, we used a Python script to manually sum all the counts for each subpixel in the matched two-dimensional image frame.

### Radio observations

Where possible, we utilized radio imaging of resolution comparable to that of Chandra (0.2–0.7″). Most radio images (32) were newly reduced for this work from archival Very Large Array, Atacama Large Millimeter/Sub-Millimeter Array (ALMA) or Australia Telescope Compact Array observations (26, 3 and 3 sources, respectively). For these we followed a standard procedure of calibrating the visibilities using the Common Astronomy Software Applications[37] software before deconvolving the image using the Common Astronomy Software Applications task clean with a Briggs (weighting) parameter of 0.5. Several rounds of self-calibration were generally used to improve the image quality. For the remaining 22 sources we used pre-existing publicly available radio images from the NASA Extragalactic Database (https://ned.ipac.caltech.edu/) (4 sources), the XJET catalogue (5 sources) or the National Radio Astronomy Observatory (NRAO) Very Large Array Archive Survey (13 sources). A complete table of the radio observations used can be found in Supplementary Table 1. Contour lines (showing levels of emission) corresponding to the radio imaging were overlaid on our co-added X-ray image as shown in Fig. 1, and we identified all distinct knots and hotspots for analysis. We ignored radio knots that were within and visually indistinguishable from the X-ray core and those that had no visible X-ray emission. We note that, while it is now well established that radio and X-ray knots do not always perfectly coincide[1], the inferred offsets from radio to X-ray are small, with most being <0.2″. This is smaller than the Chandra/ACIS pixel scale and much smaller than our region sizes.

### X-ray spectral analysis and measurement

We defined polygon regions for each knot or hotspot, being careful to avoid overlap and to encompass fully each individual emission region. For knots close to the bright central X-ray core, we used an annulus region to define simultaneously the knot (a slice of the annulus centred on the core) and its matched background (the remainder of the annulus). This is depicted in Supplementary Fig. 1 for source PKS 1928+738. In cases where a readout streak was present, this was excised proportionately from the annulus background. For knots and hotspots far from the core, a simple polygon encompassing the feature was used to extract the source counts, and a large representative region far from any source was used to estimate the background. For the 155 total jet features analysed, 93 are annulus-type regions. For jets with more than one knot region, we also defined a region encompassing the entire jet for spectral analysis. For each observation, we extracted the counts using the CIAO specextract command from every region and its associated background, including the total jet region.

To estimate the total jet spectral index we used sherpa to fit the counts extracted from the total jet region with an absorbed power-law model using the galactic hydrogen column density ($n_H$) values from the HI4PI survey[38]. In the common cases of relatively low background, it was simply subtracted. For annulus-type regions we used a second absorbed power law to model the background (dominated by the core). In a few cases, a thermal component in the background model was required to improve the spectral fit. These details are noted in Supplementary Table 1.

The total photon counts (per epoch) in our knot regions ranged from 1 to 2,156, with a median value of 38. In most cases, the photon statistics for individual knots are too low to allow us to effectively constrain the X-ray spectral index ($\Gamma_x$) for that region alone.

We attempted a spectral fit for each of the 155 regions in our sample using sherpa. Of the 52 sources in our sample, 11 sources have a single knot region of interest and thus no distinction between the spectral fit of the knot and the overall jet. The remaining 41 sources each have knot regions of interest within the jet, totalling up to 144 regions. In these cases, we compared the spectral fits we obtained from fitting the emission from all the knots together (a 'total jet' region) with the spectral fits for each individual knot. Of these, we find that the total jet $\Gamma_x$ is within $1\sigma$ of the knot value of $\Gamma_x$ for 110 regions and is within $2\sigma$ for 21 regions. The remaining 13 knot regions either have poor spectral fits (reduced $\chi^2 > 1$) or do not have enough counts for the fit to converge. Given this situation, we used the $\Gamma_x$ value from the fit to the entire jet region when generating the exposure maps for each source. We also note here the results of separate combined spectral fits to the total jet spectrum of the 13 'variable' and 40 'steady' jets, as determined by any region in the jet having single-region $P <$ or $> 0.05$, respectively. We find only a slight difference in photon index between the two samples, with the steady sources having $\Gamma_x = 1.86 \pm 0.03$ and variable sources having $\Gamma_x = 1.94 \pm 0.03$.

We used the values of $\Gamma_x$ in the CIAO task make_instmap_weights to create a weights file for each observing epoch over our energy range of 0.4–8 keV. We then used the fluximage command to generate exposure maps. We used a Python script to extract the raw counts and total exposures from our previously generated reprojected image and exposure map, respectively, for both the knot region and its associated background at every epoch.

Recent work has shown that X-ray emission from jet knots is unresolved in Chandra observations, and we accordingly treat them as point sources[1]. Since the source regions are generally small (of the order of arcseconds), the measured number of photons from the source will be less than the total falling on the detector due to the limited size of the region. The encircled count fraction is simply the ratio of the counts falling within the finite region compared with the total over the full detector. While the encircled count fraction is generally similar across epochs, it does have slightly different values due to changes in the PSF and/or changes in the detector position of the feature. We have thus used simulated PSFs from the ATLAS-X catalogue of X-ray jets and our region files to measure the encircled count fraction in each epoch. These values are tabulated in Supplementary Table 5.

### Statistical analysis method

Let $(x_i, y_i)$, $i = 1, \ldots, n$, be the background and source measurements, respectively, for a given knot, where $i$ iterates over all exposures (epochs) and $(a_i, b_i)$ are the exposures (as measured in time multiplied by area) for the background and source measurements, respectively. In addition, the source region will have an associated encircled count fraction $f_i$. The encircled count fraction and exposures are assumed to have negligible measurement error. Consider the following joint model for the observations $x_i$ and $y_i$:

$$x_i \approx \text{Poisson}(a_i \beta_i), \quad y_i \approx \text{Poisson}(b_i \beta_i + b_i f_i \mu_i), \quad (1)$$

where $\beta_i$ is the rate of background activity and $\mu_i$ is the rate of the source radiation for the $i$th observation. We are interesting in testing the null hypothesis $H_0: \mu_1 = \cdots = \mu_n$, the hypothesis that the source rates do not vary. The log-likelihood function is

$$l(\mu, \beta; X, Y) = \sum_{i=1}^{n} \{x_i \log(a_i \beta_i) - a_i \beta_i + y_i \log(b_i \beta_i + b_i f_i \mu_i) - b_i \beta_i - b_i f_i \mu_i\}, \quad (2)$$

where $\mu$, $\beta$, $X$, $Y$ are the vectorized versions of the parameters and observations.

In this section, we provide a brief discussion of the theoretical details, as well as the definitions of the necessary components for the construction of the directional test statistic used here, which has been found to be highly accurate in the small-sample setting. The full definitions and a thorough discussion of the theoretical details can be found in a previous publication[39], in which several experiments demonstrate the numerical accuracy of this method with small sample sizes. Following the general notation in ref. [39], we denote the initial parameterization by $\boldsymbol{\theta} = (\boldsymbol{\psi}, \boldsymbol{\lambda})$ where $\boldsymbol{\psi}$ is a $d$-vector parameter of interest and $\boldsymbol{\lambda}$ denotes a $(p - d)$-vector of nuisance parameters. In this model, the number of parameters increases with the number of observations as each knot requires its own source and background rate; thus, some of the expected asymptotic behaviour of the usual statistical tests may be violated. For example, we observe that the differences between competing statistical tests may not converge to 0 on the $P$-value scale, further reinforcing the need for an accurate small-sample approximation.

The $P$ value is obtained through a three-step procedure. First we reduce the dimension of the data to the number of parameters, $p$. The second step is to then further reduce the dimension to the number of parameters of interest, $d$. The third step is to then construct a one-dimensional conditional density, conditional on the direction of departure indicated by the observed data. The $P$ value is obtained by integrating this conditional density over the relevant region.

The first step is accomplished by constructing a tangent exponential model from the original model through pivotal quantities whose distributions are known[40]. This results in a density of the form

$$f_{\text{TEM}}(\mathbf{s}; \boldsymbol{\varphi}) = \exp[l\{\boldsymbol{\varphi}(\boldsymbol{\theta})\}^T \mathbf{s}(\mathbf{Y}) + l^0\{\boldsymbol{\theta}(\boldsymbol{\varphi})\}]g(\mathbf{s}), \quad (3)$$

where $l^0\{\boldsymbol{\theta}(\boldsymbol{\varphi})\}$ is the original likelihood function of the model, $\mathbf{s}$ is a constructed sufficient statistic, and $\boldsymbol{\varphi}$ is the new canonical parameterization in the tangent exponential family. A saddlepoint approximation is then used to approximate the density of this tangent exponential model, resulting in

$$f_{\text{TEM}}(\mathbf{s}; \boldsymbol{\varphi}) = \frac{\exp(k/n)}{(2\pi)^{p/2}} \exp[l(\boldsymbol{\varphi}; \mathbf{s}) - l(\hat{\boldsymbol{\varphi}}(\mathbf{s}); \mathbf{s})]|J_{\boldsymbol{\varphi\varphi}}\{\hat{\boldsymbol{\varphi}}(\mathbf{s})\}|^{-1/2}, \quad (4)$$

where $\exp(k/n)$ is the normalizing constant, $l(\boldsymbol{\varphi}, \mathbf{s})$ is the likelihood of the tangent exponential model (3)

$$l(\boldsymbol{\varphi}; \mathbf{s}) = \boldsymbol{\varphi}(\boldsymbol{\theta})^T \mathbf{s} + l^0\{\boldsymbol{\theta}(\boldsymbol{\varphi})\}, \quad (5)$$

and $J_{\boldsymbol{\varphi\varphi}}$ is the observed information matrix under the canonical parameterization of the tangent exponential family.

The second step is carried out by restricting the approximation to the plane defined by the constrained maximum-likelihood estimate $L_{\boldsymbol{\varphi}}^0 = \{\mathbf{s} : \hat{\boldsymbol{\lambda}}_{\boldsymbol{\psi}} = \hat{\boldsymbol{\lambda}}_{\boldsymbol{\psi}}^0\}$, where $\hat{\boldsymbol{\lambda}}_{\boldsymbol{\psi}}^0$ is the constrained maximum-likelihood estimate.

The saddlepoint approximation to the density for this reduced model is

$$h(\mathbf{s}; \boldsymbol{\psi}) = \frac{\exp(k'/n)^{d/n}}{2\pi} \exp\{l(\hat{\boldsymbol{\varphi}}_{\boldsymbol{\psi}}; \mathbf{s}) - l(\hat{\boldsymbol{\varphi}}; \mathbf{s})\}|\hat{j}_{\boldsymbol{\varphi\varphi}}|^{-1/2}|\tilde{j}_{(\lambda\lambda)}|^{1/2}, \quad \mathbf{s} \in L_{\boldsymbol{\varphi}}^0, \quad (6)$$

where

$$|\hat{j}_{\boldsymbol{\varphi\varphi}}| = \left|-\frac{\partial^2 l(\boldsymbol{\varphi}; \mathbf{s})}{\partial \boldsymbol{\varphi} \, \partial \boldsymbol{\varphi}^T}\right|_{\hat{\boldsymbol{\varphi}}(\mathbf{s})} \quad (7)$$

and $\tilde{j}_{(\lambda\lambda)}$ denotes the observed information matrix for the nuisance parameters, calibrated to the $\boldsymbol{\varphi}$ parameterization:

$$|\tilde{j}_{(\lambda\lambda)}| = |j_{\lambda\lambda}(\hat{\boldsymbol{\varphi}}_{\boldsymbol{\psi}}; \mathbf{s})|\left|\frac{\partial \boldsymbol{\varphi}(\boldsymbol{\theta})}{\partial \lambda}\right|_{\hat{\boldsymbol{\theta}}_{\boldsymbol{\psi}}}^{-2}, \quad (8)$$

where $\hat{\boldsymbol{\theta}}_{\boldsymbol{\psi}}$ is the constrained maximum-likelihood estimate under the $\boldsymbol{\theta}$ parameterization and $\hat{\boldsymbol{\varphi}}_{\boldsymbol{\psi}}$ is the constrained maximum likelihood under the $\boldsymbol{\varphi}$ parameterization. For the third step, following the procedure outlined in ref. [41], the one-dimensional density of $\|\mathbf{s}\|$ conditional on the direction $\mathbf{s}/\|\mathbf{s}\|$ gives the $P$ value as the ratio of integrals

$$P(\boldsymbol{\psi}) = \frac{\int_1^{t_{\max}} t^{d-1} h\{\mathbf{s}(t); \boldsymbol{\psi}\} dt}{\int_0^{t_{\max}} t^{d-1} h\{\mathbf{s}(t); \boldsymbol{\psi}\} dt}, \quad (9)$$

where $\mathbf{s}(t) = (1 - t)\mathbf{s}$ indexes points along the line, $t_{\max}$ is the maximum value of $t$ for which the constrained maximum-likelihood estimate exists and $t^{d-1}$ is the Jacobian of the transformation from $\mathbf{s}$ to $(\|\mathbf{s}\|, \mathbf{s}/\|\mathbf{s}\|)$. For more details, please see ref. [41].

Now we specialize the above discussion to the Poisson model used here. The parameters of interest are $\psi_i = \mu_{i+1} - \mu_i$ for $i = 1, \dots, n-1$, and the nuisance parameters are $\mu_1, \beta_1, \dots, \beta_n$. The null hypothesis is $H_0$: $\psi_1 = \cdots = \psi_{n-1} = 0$. The unconstrained maximum-likelihood estimators (MLEs) for $\mu_i$ and $\beta_i$ are

$$\hat{\beta}_i = \frac{x_i}{a_i}, \quad \hat{\mu}_i = \frac{y_i}{f_i b_i} - \frac{\hat{\beta}_i}{f_i}, \quad i = 1, \dots, n. \quad (10)$$

The constrained MLEs under $H_0$, $(\hat{\mu}_0, \hat{\beta}_{0,1}, \dots, \hat{\beta}_{0,n})$, are obtained by numerically maximizing

$$l(\mu^\star, \boldsymbol{\beta}; \mathbf{X}, \mathbf{Y}) = \sum_{i=1}^n \{x_i \log(a_i \beta_i) - a_i \beta_i + y_i \log(b_i \beta_i + f_i b_i \mu^\star) - b_i \beta_i - f_i b_i \mu^\star\}, \quad (11)$$

where $\mu^\star$ is scalar.

The tangent exponential family approximation to the likelihood along the line $\mathbf{s}(t)$ is

$$l\{\boldsymbol{\varphi}(\boldsymbol{\theta}); \mathbf{s}(t)\} =$$
$$\sum_{i=1}^n \left[ -(1-t)\varphi_i\{x_i - a_i \exp(\hat{\varphi}_{0,i})\} - (1-t)\varphi_{n+i}\{y_i - b_i \exp(\hat{\varphi}_{0,n+i})\} \right. \quad (12)$$
$$\left. + x_i\{\log(a_i) + \varphi_i\} - a_i \exp(\varphi_i) + y_i\{\log(b_i) + \varphi_{n+i}\} - b_i \exp(\varphi_{n+i}) \right],$$

where $\boldsymbol{\varphi}$ are the canonical parameters in the Poisson model:

$$\boldsymbol{\varphi}(\boldsymbol{\mu}, \boldsymbol{\beta}) = \begin{cases} \log(\beta_i) & \text{for } i = 1, \dots, n, \\ \log(f_{i-n}\mu_{i-n} + \beta_{i-n}) & \text{for } i = n+1, \dots, 2n. \end{cases} \quad (13)$$

The line $\mathbf{s}(t)$ linking the score at the MLE ($\hat{\boldsymbol{\varphi}}$) and the score at the constrained MLE under the null hypothesis ($\hat{\boldsymbol{\varphi}}_0$) is

$$\mathbf{s}(t) = \begin{cases} -(1-t)(x_i - a_i \exp(\hat{\varphi}_{0,i})) & \text{for } i = 1, \dots, n \\ -(1-t)(y_{i-n} - b_{i-n} \exp(\hat{\varphi}_{0,i})) & \text{for } i = n+1, \dots, 2n. \end{cases} \quad (14)$$

The $P$ value for testing $H_0$ on the basis of observations $(\mathbf{X}, \mathbf{Y})$ is obtained from the following ratio of integrals:

$$\frac{\int_1^{t_{\max}} h\{\mathbf{s}(t); \boldsymbol{\psi}\} dt}{\int_0^{t_{\max}} h\{\mathbf{s}(t); \boldsymbol{\psi}\} dt}, \quad (15)$$

where

$$h_0(\mathbf{s}(t)) = c \exp\{l(\hat{\boldsymbol{\varphi}}_0; \mathbf{s}(t)) - l(\hat{\boldsymbol{\varphi}}(\mathbf{s}(t)); \mathbf{s}(t))\} |\hat{j}_{\boldsymbol{\varphi\varphi}}|^{-1/2} |\tilde{j}_{(\lambda\lambda)}|^{1/2}. \quad (16)$$

In the original parameterization, the MLEs for the tangent exponential family approximation, given in equation (12), for a value of $\mathbf{s} = \mathbf{s}(t)$, are

$$\hat{\beta}_i(t) = t x_i / a_i + (1-t)\hat{\beta}_{0,i}, \quad (17)$$

$$\hat{\mu}_i(t) = ty_i/f_ib_i + (1-t)(\hat{\beta}_{0,i} + f_i\hat{\mu}_{0,i})/f_i - \hat{\beta}_i(t)/f_i. \quad (18)$$

As these estimates need to be positive, the maximum value that $t$ can take is

$$t_{\max} = \min\left[\min_{i \in D_1}\left\{\frac{f_i\hat{\mu}_{0,i}}{f_i\hat{\mu}_{0,i} - y_i/b_i + x_i/a_i}\right\}, \min_{j \in D_2}\left\{\frac{\hat{\beta}_{0,j}}{\hat{\beta}_{0,j} - x_j/a_j}\right\}\right], \quad (19)$$

where $D_1 := \{i \in (1, \ldots, n) : \hat{\mu}_{0,i}f_i - y_i/b_i + x_i/a_i > 0\}$ and $D_2 := \{j \in (1, \ldots, n) : \hat{\beta}_{0,j} - x_j/a_j > 0\}$.

Finally, the information terms are given by

$$|\tilde{J}_{\Phi\Phi}| = \prod_{i=1}^{n} a_i b_i \hat{\beta}_i(t)(f_i\hat{\mu}_i(t) + \hat{\beta}_i(t)), \quad (20)$$

$$|\tilde{J}_{\lambda\lambda}| = f_1^2 \left\{\frac{ty_1 + (1-t)b_1(\hat{\beta}_{0,1} + f_i\hat{\mu}_{0,1})}{(f_i\hat{\mu}_{0,1} + \hat{\beta}_{0,1})^2}\right\} \left\{\frac{tx_1 + (1-t)a_1\hat{\beta}_{0,1}}{\hat{\beta}_{0,1}^2}\right\}$$
$$\times \prod_{i=2}^{n}\left\{\frac{ty_i + (1-t)b_i(\hat{\beta}_{0,i} + f_i\hat{\mu}_{0,i})}{(f_i\hat{\mu}_{0,i} + \hat{\beta}_{0,i})^2} + \frac{tx_i + (1-t)a_i\hat{\beta}_{0,i}}{\hat{\beta}_{0,i}^2}\right\}, \quad (21)$$

where $\hat{\beta}_i(t)$ and $\hat{\mu}_i(t)$ are defined in equations (17) and (18). The $P$ value is obtained by numerically integrating the numerator and denominator of equation (15). In our simulation studies, we observe that the test is accurate for $n \leq 16$; however, beyond this point the test begins to be anticonservative under the null.

We provide some numerical simulations in Supplementary Figs. 2–4 to show that the method works well under the null for the observed configurations. We include simulations with $n = 2$, 6 and 12, with configurations described in the figure captions. We observe that the uniformity of the $P$ values is maintained.

## Data availability
All observations used in this study are publicly available. In particular, the Chandra X-Ray Observatory archive can be accessed on the web at https://cda.harvard.edu/chaser/. Radio observations with NRAO facilities are available at https://data.nrao.edu, ACTA observations from https://atoa.atnf.csiro.au/. Extensive tables of reduced data necessary to repeat these analyses are available as supplementary Excel files, further described in Supplementary Section 3.g

## Code availability
The analysis software used in this study for reducing astronomical data is described in Methods and publicly available from the respective observatories. The statistical analysis is fully described in Methods and can be repeated using any software suitable for data analysis, such as R or Python.

## Acknowledgements
The following grant funding is acknowledged: Chandra Archival Grant 16700615 (E.T.M.), ADAP grant NNX15AE55G (E.T.M.) and NSF grant 1714380 (E.T.M.), as well as support from the Natural Sciences and Engineering Research Council of Canada, grant ID RGPIN-2020-05897 (Y.T. and N.R.). This research was made possible through use of data obtained from the Chandra Data Archive and the Chandra Source Catalog, and software provided by the Chandra X-Ray Center (CXC) in the application packages CIAO and sherpa. The Australia Telescope Compact Array is part of the Australia Telescope National Facility, which is funded by the Australian Government for operation as a national facility managed by CSIRO. We acknowledge the Gomeroi people as the traditional owners of the observatory site. This paper makes use of the following ALMA data sets: 2012.1.00688.S, 2016.1.01481.S. ALMA is a partnership of ESO (representing its member states), NSF (USA) and NINS (Japan), together with NRC (Canada), MOST and ASIAA (Taiwan), and KASI (Republic of Korea), in cooperation with the Republic of Chile. The Joint ALMA Observatory is operated by ESO, AUI/NRAO and NAOJ. The NRAO is a facility of the National Science Foundation operated under cooperative agreement by Associated Universities, Inc.

## Author contributions
E.T.M. conceived the project, developed the data analysis methods, contributed to the multiwavelength data analysis and wrote the paper. A.S. carried out the X-ray data analysis with contributions from E.T.M., K.R., P.B., N.D., D.C. and M.T. Y.T. and N.R. contributed the statistical analysis methods and expertise. K.R. and P.B. contributed to the data analysis and interpretation. M.G. contributed to the interpretation and theoretical implications. All authors contributed to the editing of the paper and interpretation of results.

## Competing interests
The authors declare no competing interests.

## Additional information
**Extended data** is available for this paper at https://doi.org/10.1038/s41550-023-01983-1.

**Supplementary information** The online version contains supplementary material available at https://doi.org/10.1038/s41550-023-01983-1.

**Correspondence and requests for materials** should be addressed to Eileen T. Meyer.

**Peer review information** *Nature Astronomy* thanks the anonymous reviewers for their contribution to the peer review of this work.

**Reprints and permissions information** is available at www.nature.com/reprints.

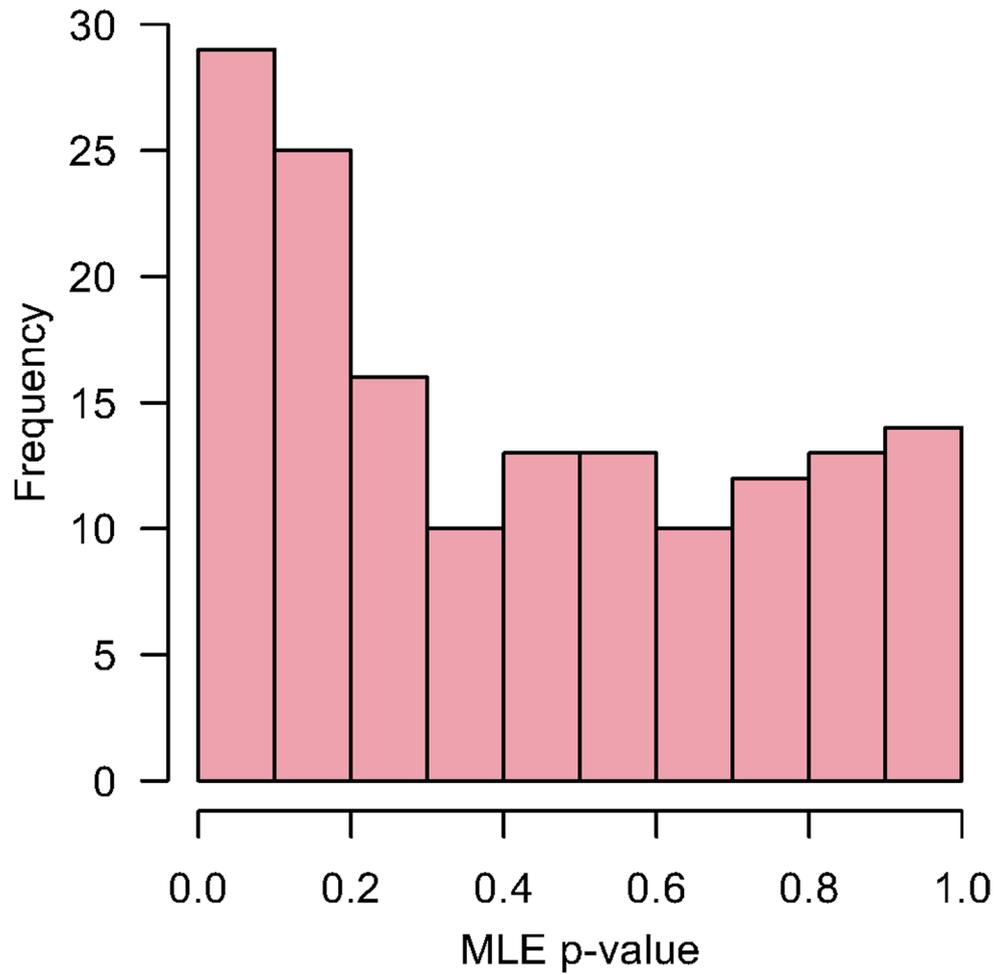

**Extended Data Fig. 1 | Histogram of the single-region p-values for the full sample of 155 regions.** The distribution shows an excess at low values relative to the expected uniform (0, 1) distribution, indicating variability in the population.

 

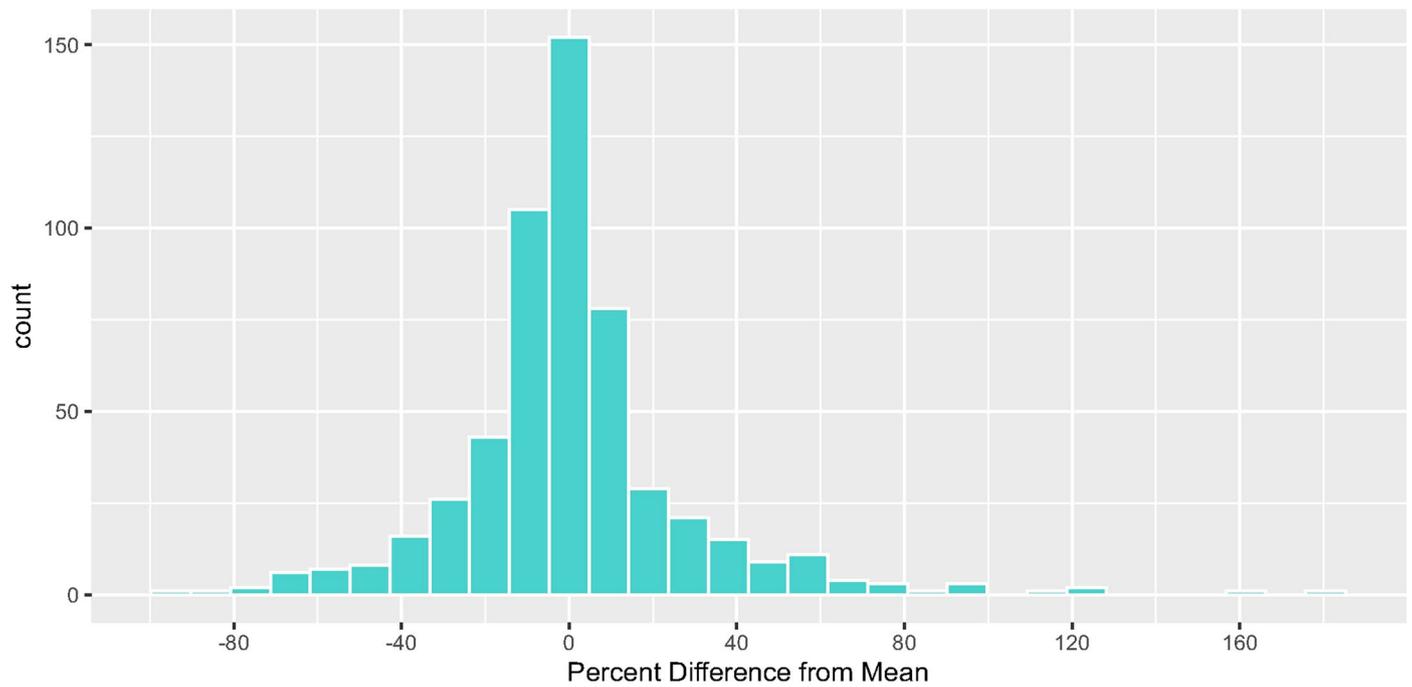

**Extended Data Fig. 2 | Histogram of the percent difference of each epoch source rate from $\bar{\mu}$ for all regions.** Including all epochs on all regions, there are 545 distinct observations. The distribution has a mean of 1.02% and a standard deviation of 28.5%. The mean of the absolute value of the percent difference is 18%.